\documentclass{iopart}

\usepackage{amssymb}

\usepackage{amsthm}

\newtheorem{Th}{Theorem}
\newtheorem{Rem}{Remark}
\newtheorem{Def}{Definition}
\newtheorem{Co}{Corollary}

\def\be{\begin{equation}}
\def\ee{\end{equation}}
\def\bea{\begin{eqnarray}}
\def\eea{\end{eqnarray}}

\def\p{\partial}
\def\s{\sigma}
\def\a{\alpha}

\def\e{\varepsilon}

\def\l{\lambda}

\def\vfi{\varphi}

\def\wt{\widetilde}

\def\cD{{\cal D}}

\def\l2{{$L^2(-\pi ,\pi )$ }}

\newcommand{\Sc}{Schr\"odinger }
\def\l2{{L^2(R^+)}}

\begin{document}

 \large

\title {Spectral singularities of non-Hermitian Hamiltonians
and SUSY transformations}

\author{
 Boris F Samsonov
}

\address{Department of Physics, Tomsk State
 University, 36 Lenin Avenue, 634050 Tomsk, Russia}

\begin{abstract}
Simple examples of non-Hermitian
 Hamiltonians with purely real spectra defined in
$L^2(R^+)$ having spectral singularities inside the continuous
spectrum  are given. It is shown that such Hamiltonians may appear
by shifting the independent variable of a real potential into the
complex plane.
Also they may be created as SUSY partners of
Hermitian Hamiltonians.
In the latter case spectral singularities of a non-Hermitian
Hamiltonian
 are ordinary points
of the continuous spectrum for its Hermitian SUSY partner.
Conditions for transformation functions are formulated when a
complex potential with complex eigenenergies and spectral
singularities has a SUSY partner with a real spectrum without
spectral singularities.
Finally we shortly discuss why Hamiltonians with spectral
singularities are `bad'.
\end{abstract}

 \submitto{\JPA}

\medskip
\medskip


\noindent

\medskip
\medskip


There are two essential differences between
 non-Hermitian Hamiltonians having a purely real spectrum and
Hermitian Hamiltonians.
The first
difference is related to the discrete spectrum
 and consists in
the possibility of appearance of {\em associated functions}
\cite{Marchenko,Naimark}
(also called in \cite{1} `background eigenfunctions'),
which
are not eigenfunctions of the Hamiltonian but they should be added
to the set of discrete spectrum eigenfunctions to complete a basis
in corresponding Hilbert space.
Hamiltonians of this kind are
known as non-diagonalizable.
In my previous Letter \cite{1} it
was shown that they can be transformed into
diagonalizable forms by appropriate {\em SUSY transformations}.
In this
note I show that SUSY transformations may be useful
to `cure' another `disease' of non-Hermitian Hamiltonians which
is related just to the second difference between Hermitian and
non-Hermitian Hamiltonians consisting in appearance of {\em
spectral singularities}
inside a continuous spectrum.

The paper by Naimark \cite{NaimarkDAN} was one of the first most
essential contribution to the spectral theory of a
non-selfadjoint operator of Schr\"odinger type.
In particular, he was the first who noticed the possibility of
appearance of spectral singularities inside a continuous
spectrum.
Later Lyantse \cite{Lyantse} (see also \cite{Naimark})
studied carefully some properties of Hamiltonians with
spectral singularities.
In this note
using the simplest real nontrivial potential without a discrete
 spectrum we demonstrate that the usual practice
of getting exactly solvable complex potentials with a purely real
spectrum consisting in simple shifting of the dependent variable
of a real potential
to the complex plane may lead to a potential with spectral
singularities.
Then we show how a spectral singularity may appear
 after a SUSY transformation over a real potential
and how it can be `removed'
(more precisely it can be transformed into an ordinary point of
the continuous spectrum of a SUSY partner Hamiltonian).
Finally we shortly discuss why
Hamiltonians possessing spectral singularities are `bad'.

For simplicity we will consider Sturm-Liouville problems on the
positive semiaxis since in this case the continuous spectrum is
non-degenerate and spectral singularities are easier `to handle'.
Nevertheless, we would like to point out that interested reader
can find a deep
study of non-Hermitian Hamiltonians defined on the whole real
line,
which includes the strict formulation and proof of the inverse
scattering theorem, in \cite{Blaschak}. (We notice that spectral
singularities constitutes a part of spectral data.)
We start with the definition of spectral singularities
 of a non-Hermitian
one-dimensional Hamiltonian
 mainly following the paper by Lyantse \cite{Lyantse}.

Let us have a complex-valued function $V(x)$ such that
\be\label{condA}
\int_0^\infty e^{\e x}|V(x)|dx <\infty
\ee
where $\e$ is a positive number.
We call such functions {\em exponentially decreasing} at the infinity.
We note that this condition simplifies essentially studying
non-Hermitian Hamiltonians with spectral singularities although
they may appear under a weaker condition that $V(x)$ is absolutely
summable on the positive semiaxis.
In particular, condition (\ref{condA}) guaranties a finite number of
spectral singularities.

Consider the differential expression
\be\label{de}
l[y]=-y''+V(x)y
\ee
and the boundary condition
\be\label{bc}
y(0)=0\,.
\ee
Let $h$ be (non-selfadjoint)
 operator created by differential expression (\ref{de})
and condition (\ref{bc}) in the Hilbert space $\l2$
where $R^+$ is the positive semiaxis $R^+=(0,\infty)$.
The domain of definition of $h$, $\cD(h)$, consists of all
functions $\psi$, which have absolutely continuous derivative in any
finite interval $(\alpha,\beta)\subset R^+$, satisfy the
boundary condition (\ref{bc}) and such that both $\psi\in\l2$ and
$l[\psi]\in\l2$. If $\psi\in\cD(h)$ then by definition
$h\psi=l[\psi]$,
so, we also denote $h=-\p_x^2+V(x)$ and call $h$ Hamiltonian.

For any potential $V(x)$ satisfying condition (\ref{condA}) there
exists a solution $e_+(x,s)$ (Jost solution)
 of the equation $l[y]=s^2y$ with the following
asymptotical behavior at $x\to\infty$:
\be
\frac{d^j}{dx^j}e_+(x,s)=(ix)^je^{isx}+o(e^{-\frac{\e}{2}x})
\qquad j=0,1,\ldots
\ee
which is uniform with respect to $s$ in any half-plane
$\mbox{Im}s\geqslant-\eta$ where $0<\eta<\frac{\e}{2}$.
Put
\be\label{As}
A(s)=e_+(0,s)\,.
\ee
The function $A(s)$ (Jost function) is holomorphic in the
half-plane
$\mbox{Im}s>-\frac{\e}{2}$ and its asymptotics
\be
A(s)=is[1+o(1)]\qquad |s|\to \infty
\ee
implies that the equation $A(s)=0$ has a finite number of
solutions in the half-plane $\mbox{Im}s\geqslant 0$.
Let $s_1,\ldots, s_r$ be solutions to the equation $A(s)=0$ in the
open upper half-plane $\mbox{Im}s> 0$. Then the spectrum of the
operator $h$ has the discrete part defining as
$E_1=s_1^2,\ldots, E_r=s_r^2$ and the continuous part filling the
non-negative semiaxis $E\geqslant0$.

For a selfadjoint Hamiltonian the Jost function has no real zeros
\cite{Levitan}.
But for a non-selfadjoint case
it may happen that the equation $A(s)=0$ has real roots.
Let
$\s_1,\ldots,\s_\rho$ be real non-zero solutions to this equation.
We would like to stress that they may be both positive and
negative.
Since $\s_1^2>0,\ldots ,\s_\rho^2>0$, the points
\be\label{singP}
 \wt E_1=\s_1^2,\ldots ,\wt E_\rho=\s_\rho^2
\ee
 belong to the continuous
 spectrum of $h$.

\begin{Def}
Points (\ref{singP}) are called {\it spectral singularities} of the
operator $h$.
\end{Def}

\begin{Rem}
 We would like to stress the difference between
spectral singularities and other exceptional points that may
appear inside the continuous spectrum which are known as bound
states embedded into continuum
(BEICs, see e.g. \cite{Stahl} and references therein).
In contrast with spectral
singularities, which belong to the continuous spectrum, BEICs
correspond to the discrete spectrum of a Hamiltonian.
\end{Rem}

Inside the strip $|\mbox{Im}s|\leqslant\frac {\e}{2}$ the functions
$e_+(x,s)$ and $\wt e_-(x,s)=e_+(x,-s)$ form a fundamental set of
solutions to the equation $l[y]=s^2y$ and their Wronskian is
$W(e_+,\wt e_-):=e_+(x,s)\wt e\hspace{.2em}'_-(x,s)-
e'_+(x,s)\wt e_-(x,s)=-2is$ where the
prime denotes the derivative with respect to $x$.
From here it follows that the function defined
as
\be
\psi(x,s^2)=
\frac{A(-s)e_+(x,s)-A(s)e_+(x,-s)}{-2is} \qquad
|\mbox{Im}s|\leqslant \frac{\e}{2}
\ee
satisfies the condition $\psi(0,s^2)=0$ and  $\psi'(0,s^2)=1$.

For any $s$ from the half-plane $\mbox{Im}s>0$ there exists a function
$e_-(x,s)$ exponentially growing as $x\to\infty$ which
 in this case is
 a solution linearly independent with $e_+(x,s)$ and
 the Wronskian of $e_+$ and $e_-$ has the same value as in the previous case,
$W(e_+,e_-)=-2is$. Therefore we can also write
\be
\psi(x,s^2)=
\frac{B(s)e_+(x,s)-A(s)e_-(x,s)}{-2is} \qquad
\mbox{Im}s> 0
\ee
where $B(s)=e_-(0,s)$.
Keeping in mind the case of a self-adjoint operator $h$ we can say
that
at $s^2=E_1,\ldots,s^2=E_r$ $\psi(x,s^2)$ describes
(unnormalized) bound states of the
hamiltonian $h$ and for $\mbox{Im}s=0$ this is a scattering state.
We would like to stress that the physical contents of a scattering
of a particle on a complex potential needs a special analysis which
is out of the scope of the current note and will be discussed in a
separate publication.


Let us consider the well-known exactly solvable potential
\be\label{1soliton}
V(x)=\frac{-2 a^2}{\cosh^2(ax)}\qquad a>0 \,.
\ee
Being considered on the whole real line this is a transparent
one-level (one-soliton) potential but on the positive semiaxis it
has only a continuous spectrum.
Let us shift the space variable to the complex plain and
consider
\be\label{example1}
V(x)=\frac{-2 a^2}{\cosh^2(ax+b)}\qquad 0\leqslant x<\infty
\qquad b\in\Bbb C\qquad a>0\,.
\ee
This potential evidently satisfies the condition
(\ref{condA})
for any $\e<2a$.

It is not difficult to find the Jost solution
\be\label{fik0}
e_+(x,k)=\frac{k+ia\tanh (b+ax)}{k+ia}\,e^{ikx}
\ee
which according to (\ref{As})
gives the Jost function
\be\label{Ak}
A(k)=\frac{k+ia\tanh (b)}{k+ia}\,.
\ee
Here and in what follows we are using the notation
$k=s$ for Im$s=0$.
The function (\ref{Ak}) vanishes at $k=k_0=-ia\tanh b$ which is real
provided $b$ is purely imaginary.
Thus, for a purely imaginary $b$ potential (\ref{1soliton})
 at $k=k_0=-ia\tanh b$
 has a
spectral singularity $E=k_0^2$.
The position of this point does not depend
 on the sign of
$k_0$ and in particular on the sign of
  the parameter $b$ but
the shape of the potential (\ref{example1})
changes with the change of this
 sign.

Now we are going to show how similar Hamiltonians occur after
SUSY transformations over Hermitian Hamiltonians. Since such
transformations can always be realized in the opposite direction this
opens the possibility to convert
(non-Hermitian) Hamiltonians with spectral singularities
into Hermitian Hamiltonians.


Let us denote $V=V_0$ in (\ref{de}).
We will suppose that $V_0(x)$ is a real scattering potential
(see e.g. \cite{Levitan}) i.e. it satisfies the
condition
 \be\label{cond}
\int_0^\infty x|V_0(x)|dx<\infty\,.
 \ee
 In this case
 the operator $h=h_0=-\p_x^2+V_0(x)$
 is essentially self-adjoint
in the Hilbert space $L^2(R^+)$
 and
 has a real and simple
spectrum
which has a finite number (which may be equal to zero)
 of discrete levels and a continuous part filling the non-negative
 semiaxis \cite{Levitan}. We will denote $\psi_n(x)$ its discrete
 spectrum eigenfunctions and
 $\psi_k(x)$, $E=k^2>0$ the continuous spectrum eigenfunctions
 keeping the notation $\psi_E(x)$ for a solution to the differential
  equation
 $-\psi_E''(x)+[V_0(x)-E]\psi_E(x)=0$.

 To simplify notations in what follows we
 will use symbols $h_0$, $h_1$, $h_2$ to define both the
 operator-differential expressions $h_j=-\p_x^2+V_j(x)$
 $j=0,1,2$ and corresponding
 operators in the Hilbert space $\l2$
  calling them Hamiltonians.
 We hope this will not cause
 troubles for the reader.

 For getting a complex potential $V_{2}(x)$
defining the Hamiltonian $h_{2}=-\p_x^2+V_{2}(x)$,
which is a 2-SUSY partner
 to the Hermitian $h_0$,
 we are using
  {\em second
   order SUSY transformations} considered
 in necessary details in \cite{1}.
Below
 we formulate conditions which should be imposed on transformation
 functions leading to appearance of a spectral singularity in the
 spectrum of $h_2$. Accordingly, the opposite transformation
 removes the spectral singularity and gives back (Hermitian)
 Hamiltonian $h_0$.
 If the 'wave function'
(we are using physical terminology for non-Hermitian Hamiltonians
also)
 corresponding to
 the spectral singularity is nodeless it
 can be used for removing the spectral singularity by a simpler first
 order SUSY transformation.

Before formulating the main result of the present Letter
 we remind very briefly necessary formulas defining
 1- and 2-SUSY transformations.

If the first order transformation
is implemented over the potential $V_0$
 the 1-SUSY partner
potential $V_1(x)$
is given by
\be\label{V11}
V_1(x)=V_0(x)-2w'(x)\qquad w(x)=[\log u(x)]'
\ee
where $u(x)$ is a nodeless $\forall x\in R^+$
solution to the equation
$h_0u=\a u$.
It is called {\it transformation function}.
It can vanish at the origin thus giving a potential
singular at the origin.
The solutions to the transformed equation
$h_1\vfi_E=E\vfi_E$,
$h_1=-\p_x^2+V_1(x)$
are expressed in terms of solutions $\psi_E(x)$
to the initial equation
\bea
 \vfi_E(x)=-\psi'_E(x)+w(x)\psi_E(x)\qquad
 E\ne\a
 \\
\vfi_\a(x)={\frac {1}{u(x)}}\,.
 \eea
Choice $\vfi_\a={\frac 1u}$
as the transformation function
for the next 1-SUSY transformation
corresponds to the backward transformation from $h_1$ to $h_0$.

In the case of a second order transformation
over the potential $V_0$
the 2-SUSY partner
potential $V_2$
reads
\be\label{V1}
V_2=V_0-2 \left[\log W(u_1,u_2)\right]''
\ee
and the solutions to the equation $h_2\vfi_E=E\vfi_E$,
 $h_2=-\p_x^2+V_2(x)$
 are defined as
follows:
\bea\label{fi1}
&\vfi_E
=(E-\a_2)\psi_E+(\a_1-\a_2)
\frac{W(u_2,\psi_E)}{W(u_1,u_2)}\,u_1\\
&\hphantom{\vfi_E} =  (E-\a_1)\psi_E+(\a_1-\a_2)
\frac{W(u_1,\psi_E)}{W(u_1,u_2)}\,u_2
\qquad E\ne \a_1,\a_2
\label{fi2} \\
  \label{fial}
&\hspace{-.7em} \vfi_{\a_{1,2}}  =
\frac{u_{2,1}}{W(u_1,u_2)}\,.
\eea
Here $u_1$, $u_2$
(they are called {\it transformation functions})
and $\psi_E$ are solutions to the equation
$h_0y=Ey$
corresponding to the eigenvalues $\a_1$, $\a_2$  ($\a_1\ne\a_2$)
 and $E$ respectively;
 the symbol $W(\cdot,\cdot)$ is reserved for the Wronskian.
 If we realize the same transformation once again choosing
 $h_2$ as the initial Hamiltonian and the functions
 (\ref{fial}) as the transformation functions we will go back to the
 potential $V_0$. Thus, the procedure is completely reversible.

Now we are able to prove the main result of the present note.\\
\begin{Th}
Let a real-valued function $V_0(x)$, $x\in R^+$
be smooth, finite at $x=0$
and exponentially decreasing at the infinity.
Then there exists a 2-SUSY transformation
such that a complex potential $V_2(x)$ obtained with the help of
formula (\ref{V1}) is
exponentially decreasing and
regular $\forall x\in [0,\infty)$;
the Hamiltonian $h_2=-\p_x^2+V_2(x)$
has a spectral  singularity at $E=k_0^2$
where $k_0$ is an arbitrary non-zero real number.
There exists a real 1-SUSY partner $V_1(x)$ of the potential
$V_2(x)$ singular only at the origin with the singularity strength
equal $1$.
The Hamiltonian $h_1=-\p_x^2+V_1(x)$ is essentially selfadjoint in
$\l2$.
\end{Th}
\begin{Rem} We remind
(see e.g. \cite{PRC1})
that singularity strength $\nu$ of a
potential $V(x)$ is defined by its behavior at the origin:
\be\label{ss}
V(x)\to \frac{\nu(\nu+1)}{x^2}\qquad x\to0\,.
\ee
\end{Rem}
\begin{Rem}
Our proof of the theorem is constructive so that
below we give a precise
recipe how such potentials may be obtained.
\end{Rem}

{\em Proof.}
To prove the first part of the statement we formulate conditions
which should be imposed
on transformation functions $u_1$ and $u_2$ leading
 according to (\ref{V1}) to the potential $V_2$
 with desirable properties.
First we notice that $V_0(x)$ satisfies
 condition (\ref{cond}) and
any solution to the equation
$(h_0-E)\psi_E=0$
with the given $E\in \Bbb C$
 at the origin either vanishes (regular solution) or takes a
 finite non-zero value (irregular solution).
 It follows from here and (\ref{fi1},\ref{fi2})
 that to preserve the zero boundary condition at the origin the
 method should involve transformation functions $u_1(x)$ and
 $u_2(x)$ such that at least one of them vanishes at the
 origin.

 Let $u_1(x)$ be real and $u_1(0)=0$, $\a_1<0$.
  Choose $u_2(x)$
coinciding with the Jost solution $u_2(x)=e_+(x,k_0)$,
($\a_2=k_0^2>0$).
Formula (\ref{V1}) gives a regular potential $V_2(x)$ provided the
Wronskian of $u_1$ and $u_2$ is nodeless in $R^+$.
Therefore first of all we have to convince ourselves that this property
of the Wronskian takes
place for the given $u_1$ and $u_2$.

According to Sturm oscillator theorem (see e.g. \cite{BerSh}) $u_1(x)$
has no nodes in $R^+$.
The same property takes place for $u_2(x)$ also.
 Indeed, since $u_2(x)$ is a complex solution to the \Sc equation
with a real potential $V_0(x)$ it can always be presented as
$u_2(x)=y_1(x)+iy_2(x)$ where real-valued functions $y_1(x)$ and
$y_2(x)$ are two linearly independent solutions to the same
equation with the same value of $E=\a_2$. Since they are real and
correspond to $E>0$ they have an oscillating character but their
nodes never coincide.
Otherwise their Wronskian which is $x$-independent would vanish
which is impossible.

We shall now show that $u_1$ and $u_2$ have a non-vanishing
Wronskian $\forall x\in[0,\infty)$.
Denote for brevity $W=W(u_1,u_2)$
which is a smooth complex-valued function of the real variable $x$. Let
$f$ be its modulus, $f=\sqrt{WW^*}$, where the asterisk denotes the
quantity complex conjugate to the given one
and only the positive branch of the square root should be used.
Suppose that the function $W=W(x)$ vanishes at any
$x=x_0\in R^+$, $W(x_0)=0$.
Since $f(x)\geqslant0$ $\forall x\in R^+$ the point $x=x_0$
is the point of a
minimum for $f(x)$.
Then since $W^*(x_0)=0$, the ratio
$W^*/W$ is undetermined.
Using the l'Hospital rule
we find  that $W^*/W=(W^*)'/W'$ at $x=x_0$.
From here we get
\[
f'(x_0)=\frac{W'}{2}\sqrt{\frac{W^*}{W}}+
\frac{(W^*)'}{2}\sqrt{\frac{W}{W^*}}=
|{W'(x_0)}|\,.
\]
Using the \Sc equation
and already established fact that both
$u_1(x)$ and $u_2(x)$ are nodeless in $R^+$
we can easily see that
$W'(x)=(\a_2-\a_1)u_1(x)u_2(x)$ does not vanish
$\forall x\in R^+$. This result means that $f'(x_0)\ne0$
which contradicts to the fact that $x_0$ is the point of a minimum
for $f(x)$.
This contradiction proves the nodeless character of
$W(x)$ $\forall x\in R^+$.
Therefore taking into consideration that $W(0)=-u_1'(0)u_2(0)\ne0$
we see that the potential $V_2(x)$ given in (\ref{V1}) is regular
$\forall x\in [0,\infty)$.
The fact that $V_2(x)$ is exponentially
decreasing
 follows from formula (\ref{V1}) and the asymptotical behavior
of the transformation functions $u_1(x)$ and $u_2(x)$.

Now we shall prove that
 the point $E=k_0^2$ of the continuous
 spectrum of the Hamiltonian $h_2=-\p_x^2+V_2(x)$ with $V_2(x)$
 given in (\ref{V1}) is a spectral singularity.
 Indeed,
according to (\ref{fial}) the condition
 $\vfi_{\a_2}(0)=0$ follows from the regular character of
 $u_1(x)$. Necessary asymptotical behavior of $\vfi_{\a_2}(x)$ follows
 from the asymptotics of the Jost solution $e(x,k_0)$
 and the fact that the
 logarithmic derivative of $u_1(x)$ is asymptotically constant.
 Thus, this is just the point $E=\a_2=k_0^2$ which is a spectral
 singularity for $h_2$.
 We note that the choice $u_2(x)=e(x,-k_0)$ produces another potential
 with the same spectral singularity at $E=k_0^2$.

 To prove the last statement of the theorem we note that the
 second order transformation leading to $V_2$ from the given $V_0$
 can always be presented as a chain of two transformations
 $V_0\to V_1\to V_2$ where a real-valued
 potential $V_1$ is obtained from $V_0$ by the
 first order transformation (\ref{V11}) with a real transformation
 function $u=u_1(x)$ vanishing only at the origin and this zero is
 simple. Therefore $V_1(x)$ has only one singular point (poles)
 $x=0$ and its behavior near $x=0$ is given by (\ref{ss}). The
regular potential $V_2$ is obtained from $V_1$ by another first
order transformation with the transformation function
$\wt u(x)=-u_2'(x)+w(x)u_2(x)$ where $w(x)=u_1'(x)/u_1(x)$.
Therefore using the transformation function $1/\wt u(x)$
which is just the continuous spectrum eigenfunction corresponding
to the spectral singularity $E=k_0^2$
we go
back from $V_2$ to $V_1$ transforming in this way the complex
potential $V_2$
having a spectral singularity at $E=k_0^2$
 into the real potential $V_1$ without spectral singularities but
 singular at $x=0$ with the singularity strength equal 1.
 The value 1 for the singularity strength follows from the fact
 that the zero at $x=0$ of the transformation function $u_1(x)$ is
 simple.
 The Hamiltonian $h_1$ is essentially self adjoint since
 the potential $V_1(x)$
 is real, finite in $R^+$ and exponentially
 decreasing at the infinity (the 1-SUSY transformation also does not
 change the asymptotical behavior of the potential).
 \hfill $\square$

From the last lines of the proof we can draw the following
deductions.
 \begin{Co}
Suppose that the continuous
spectrum eigenfunction $\psi_{k_0}(x)$
of a regular in $R^+$ complex potential $V_2(x)$,
which is a 2-SUSY partner of a real and regular
$V_0(x)$,
corresponds to a spectral
singularity $E=k_0^2$ such that the zero $k=k_0$ of the Jost
function
$A(k)$ is simple then $\psi_{k_0}(x)$
is nodeless.
\end{Co}
More important for us is the following implication.
\begin{Co}
Let a complex potential $V_2(x)$ has only one spectral singularity
$E=k_0^2$, the zero $k=k_0$ of the Jost function $A(k)$ is simple
and continuous spectrum eigenfunction $\psi_{k_0}(x)$ is nodeless
in $R^+$.
Then $V_2(x)$,
which
 is 1-SUSY partner obtained with the help
of (\ref{V11}) with $u=\psi_{k_0}$,
 has no spectral singularities
 and is singular only at the origin with the singularity strength
 equal 1.
 \end{Co}

 For instance for $u=\vfi_{k_0}$ where
$\vfi_{k_0}$ is given in (\ref{fik0})
at $k=k_0=-ia\tanh b$
the potential
$V_1$
which is 1-SUSY partner of $V_2$ given in (\ref{example1})
reads
\[
V_1=\frac{2 a^2}{\sinh^2(ax)}\,.
\]

 The last statement
while combined with results of the paper \cite{1} leads to
the following theorem
 which we leave without proving.
 \begin{Th}
Let finite at the origin and
exponentially decreasing
complex potential $V_{N+n}(x)$ be such that the equation
$A(k)=0$
where $A(k)$ is the Jost function
has only simple roots, between which
$n$  non-zero roots $k_0,\ldots ,k_{n-1}$ are real
(they correspond to spectral singularities $E_j=k^2_j$,
$j=0,\ldots,n-1$
of the Hamiltonian $h_{N+n}=-\p_x^2+V_{N+n}$)
and $N$ roots $s_1,\ldots,s_N$
lay in the open upper half of the complex
$s$-plane outside the imaginary axis
($E_l=s_l^2$, $l=1,\ldots,N$
 are points of the discrete spectrum of $h_{N+n}$).
If the Wronskian
$W(\wt\psi_{s_1},\ldots,\wt\psi_{s_N},
\psi_{k_0},\ldots,\psi_{k_{n-1}})$,
where
$\wt\psi_{s_l}$ are discrete spectrum eigenfunctions
and
$\psi_{k_j}$, $j=0,\ldots ,n-1$ are continuous spectrum
eigenfunctions, does not vanish in $R^+$ then the potential $V_{N+n}(x)$
is a $(N+n)$-SUSY partner of the potential
\be
V_0(x)=V_{N+n}-2[\log W(\psi_{s_1},\ldots,\psi_{s_N},
\psi_{k_0},\ldots,\psi_{k_{n-1}})]''
\ee
 which has a real spectrum,
 has no spectral singularities and
  is singular at the
origin with the singularity strength equal $N+n$.
 \end{Th}

Let us consider a few examples illustrating Theorem 1.

{\em Example 1.} Take $V_0=0$,
$u_1(x)=\sinh(a_1x)$, ($\a_1=-a_1^2$), $u_2=\exp(-ik_0x)$.
Formula (\ref{V1}) gives the potential
\be\label{ex1}
V_1(x)=-\frac{2a_1^2(a_1^2+k_0^2)}{[a_1\cosh(a_1x)-ik_0\sinh(a_1x)]^2}
\ee
which
at $a_1=a$ and $k_0=ia\tanh b$
is easily identified as the one given in (\ref{example1}).
This means that the potential (\ref{example1}) having a
spectral singularity
 is 2-SUSY partner for the
zero potential having no spectral singularities.

{\em Example 2.} Take $V_0=-6\mbox{sech}^2x$.
This is a one-level potential with the ground state function
$\psi_0=\sqrt 3\tanh x\,\mbox{sech}x$ ($E_0=-1$) and the
Jost solution
\be
e(x,k)=\frac{e^{ikx}}{(k+i)(k+2i)}
(k^2 -2+3\,\mbox{sech}^2x + 3i k \tanh x)\,.
\ee
The choice $u_1=\psi_0$ and $u_2=e(x,k_0)$
leads to eliminating the ground state level from the spectrum of
$V_1$ and gives the potential (\ref{ex1}) at $a_1=2$.
Since SUSY transformations are reversible this result means that
the potential (\ref{ex1})  at $a_1=2$ is  SUSY partner for both $V_0=0$ and
 $V_0=-6\mbox{sech}^2x$.

{\em Example 3.} Take finally $V_0=-20\,\mbox{sech}^2x$.
The Hamiltonian $h_0$ has two
discrete levels: $E_0=-9$,
$\psi_0=\sqrt{105/8}\,\mbox{sech}^3x\tanh x$ and $E_1=-1$,
$\psi_1=\sqrt{5/8}[2\cosh(2x)-5]\,\mbox{sech}^3x\tanh x$.
The Jost solution has the form
\bea \nonumber
& e(x,k)=
\frac{e^{ikx}}{(k+i)(k+2i)(k+3i)(k+4i)}\,[k^4-35k^2+24 \\
&  +105\,\mbox{sech}^4x+10ik(k^2-5)\tanh x+
15\, \mbox{sech}^2x(3k^2-8+7ik\tanh x)]\,.\nonumber
\eea
Once again we choose
 $u_1=\psi_0$ and $u_2=e(x,k_0)$.
 Up to an inessential constant factor the Wronskian of these
 functions reads
 \[
W(u_1,u_2)=
e^{ik_0x}\,w_0(x)\,\mbox{sech}^7x
\]
where
$w_0(x)=
A_1\cosh2x+A_2\cosh4x+B_1\sinh2x+B_2\sinh4x +c$
with
$A_1=4(16+k_0^2)$, $A_2=7k_0^2-8$, $B_1=-2ik_0(16+k_0^2)$,
$B_2=-ik_0(k_0^2-14)$ and $c=-3(k_0^2+16)$.
Formula (\ref{V1}) yields the potential
\[
V_1=-\frac{6}{\cosh^2x}+2\,\frac{[w'_0(x)]^2-w_0''(x)w_0(x)}{w_0^2(x)}
\]
It has one discrete level $E=-1$ and
the spectral  singularity $E=k_0^2$.

Now we discuss shortly the role of spectral singularities
\cite{Naimark,Lyantse,Levin}.

(I) First of all we would like to stress that the spectral
singularities play a very special role with respect to the other
points of the continuous spectrum of the operator $h$. Hamiltonian
$h$ having spectral singularities has some new properties which
 appear neither in the theory of self-adjoint operators no in the
 theory of non-Hermitian Hamiltonians with purely discrete
 spectrum and in particular
in the theory of
 finite-dimensional Hamiltonians.

(II)
It happens that the whole Hilbert space
$L^2(R^+)$ can be presented as an
orthogonal sum ${\cal M}\dot + {\cal N}$ where $\cal M$ corresponds to
the continuous spectrum of $h$ and  $\cal N$ is related
to its discrete spectrum.
Under condition (\ref{condA}) the
space  $\cal N$ is finite-dimensional.
If $h$ has no spectral singularities
being restricted to the space  $\cal M$
 it is similar to a self-adjoint operator.
 This means that there exists a continuous
in the $L^2(R^+)$-norm
  one-to-one mapping
 of the space  $\cal M$ onto itself which we denote $T$ such that
 $T^{-1}hT$ is self-adjoint. If $h$ has a spectral
 singularity such a representation is impossible.

(III)
With the operator $h$ one can always associate so called
`resolution of the identity operator' $P=P(\Delta)$ which is a
set of projectors having properties similar to
corresponding properties in the case
of a Hermitian operator. It plays a role of a `coordinate system'
where the `matrix' of the operator $h$ takes a `simplest' form.
If $h$ has no spectral singularities the resolution of the
identity is bounded with respect to a norm. For $h$ with
spectral singularities the norm $\|P(\Delta)\|$ tends to infinity
when $\Delta$ tends to a spectral singularity. In geometric
language this may be interpreted as if  `coordinate
system' where $h$ with a spectral singularity
 acquires the simplest form has such `coordinate subspaces' that the
 angle between them may be as small as desired.

(IV)
According to a result by Levin \cite{Levin}
 for any $h$ without spectral singularities there exists
 for all $f\in L^2(R^+)$ a series expansion over the set of
 eigenfunctions and associated functions converging in
 $L^2(R^+)$-norm. If $h$ has a spectral singularity such
 an expansion exists not for all $f\in L^2(R^+)$ but the set
 for which it exists is dense in $L^2(R^+)$
 \cite{Lyantse}.
 This feature of a Hamiltonian with spectral
 singularities is due to the fact that
for some values of $k=k_j$
the Jost function vanishes, i.e.
either
$A(k_j)=0$ or $A(-k_j)=0$.
Just both $A(k)$ and $A(-k)$
 appear in the denominator of the integral over the continuous
 spectrum in
 the Fourier series expansion of an element from the Hilbert space
 $L^2(R^+)$ over the set of the eigenfunctions
 and associated functions of the Hamiltonian $h$.
 As a result at either
$k=k_j$ or $k=-k_j$
 the integrand has a pole and the integral becomes
divergent. In \cite{Lyantse} a regularization procedure for this
integral is discussed in details.

 We leave open the question whether or not
  Hamiltonians
with spectral singularities
  are acceptable in
 complex quantum mechanics. But we hope that in some
 cases SUSY transformations may be useful while working with such
 Hamiltonians.

The work is partially supported by the
President Grant of Russia 1743.2003.2
 and the Spanish MCYT and European FEDER grant BFM2002-03773.

\end{document}